\documentclass[structabstract]{aa} 

\usepackage{graphicx}
\usepackage{amsmath}
\usepackage{txfonts}
\usepackage{natbib}

\newcommand{\bo}{\ensuremath{\boldsymbol{B}_0}}
\newcommand{\dm}{\ensuremath{D_{\mu\mu}}}
\newcommand{\abs}[1]{\ensuremath{\left\lvert #1\right\rvert}}

\newcommand{\Rl}{\ensuremath{R_{\mathrm L}}}

\newcommand{\be}{\begin{equation}}
\newcommand{\ee}{\end{equation}}
\newcommand{\bs}{\begin{subequations}}
\newcommand{\es}{\end{subequations}}

\newcommand{\ez}{\ensuremath{\hat{\boldsymbol{e}}_z}}

\newcommand{\pa}{\ensuremath{_\parallel}}

\newcommand{\De}{\ensuremath{\varDelta}}
\newcommand{\Om}{\ensuremath{\varOmega}}
\newcommand{\Ph}{\ensuremath{\varPhi}}
\newcommand{\Ga}{\ensuremath{\varGamma}}

\newcommand{\usum}{\ensuremath{\sum_{n=-\infty}^\infty}}

\newcommand{\f}[1]{\ensuremath{\boldsymbol{#1}}}
\newcommand{\m}[1]{\ensuremath{\left\langle #1\right\rangle}}

\newcommand{\dd}[2][]{\ensuremath{\frac{\mathrm{d} #1}{\mathrm{d} #2}}}
\newcommand{\df}{\ensuremath{\mathrm{d}}}

\begin{document}
\title{Pitch-angle scattering in magnetostatic turbulence}
\subtitle{I. Test-particle simulations and the validity of analytical results}
\author{R.\,C. Tautz\inst{1} \and A. Dosch\inst{2} \and F. Effenberger\inst{3,4} \and H. Fichtner\inst{4} \and A. Kopp\inst{5}}
\institute{Zentrum f\"ur Astronomie und Astrophysik, Technische Universit\"at Berlin, Hardenbergstra\ss{}e 36, D-10623 Berlin, Germany\\\email{rct@gmx.eu}
\and Center for Space Plasmas and Aeronomic Research, University of Alabama in Huntsville, 320~Sparkman Drive, Huntsville, AL~35805, USA
\and Department of Mathematics, University of Waikato, PB~3105, Hamilton, New Zealand
\and Institut f\"ur Theoretische Physik, Lehrstuhl IV: Weltraum- und Astrophysik, Ruhr-Universit\"at Bochum, D-44780 Bochum, Germany
\and Institut f\"ur Experimentelle und Angewandte Physik, Christian-Albrechts-Universit\"at zu Kiel, Leibnizstra\ss{}e 11, D-24118 Kiel}

\date{Received June 26, 2013; accepted August 21, 2013}

\abstract
{Spacecraft observations have motivated the need for a refined description of the phase-space distribution function. Of particular importance is the pitch-angle diffusion coefficient that occurs in the Fokker-Planck transport equation.}
{Simulations and analytical test-particle theories are compared to verify the diffusion description of particle transport, which does not allow for non-Markovian behavior.}
{A Monte-Carlo simulation code was used to trace the trajectories of test particles moving in turbulent magnetic fields. From the ensemble average, the pitch-angle Fokker-Planck coefficient is obtained via the mean square displacement.}
{It is shown that, while excellent agreement with analytical theories can be obtained for slab turbulence, considerable deviations are found for isotropic turbulence. In addition, all Fokker-Planck coefficients tend to zero for high time values.}
{}

\keywords{Plasmas --- Magnetic Fields --- Turbulence --- (Sun:) solar wind --- (ISM:) cosmic rays}
\authorrunning{Tautz et al.}
\titlerunning{Pitch-angle scattering in magnetostatic turbulence}
\maketitle

\section{Introduction}

Recent spacecraft observations have revealed the necessity to refine the modeling of the transport of charged energetic particles to allow for strongly pitch-angle--anisotropic phase space distribution functions, which cannot be properly accounted for by the diffusion approximation. In addition to solar energetic particles (SEPs) in general, for which this requirement has been known for decades \citep[][for a review see \protect\citealt{Droege-2000}]{Roelof-1969}, other heliospheric particle populations were identified to exhibit such anisotropies. Examples are the so-called Jovian electron jets \citep{Ferrando-etal-1993, Dunzlaff-etal-2010} and suprathermal ion species accelerated at interplanetary traveling shocks \citep{leRoux-Webb-2012}, as well as at the solar wind termination shock \citep{Decker-etal-2005, Florinski-etal-2008, leRoux-Webb-2012}.

Central to such a modeling refinement is the determination of the pitch-angle diffusion coefficient $D_{\mu\mu}$ that occurs in the Fokker-Planck transport equation \citep{rs:rays,sha09:nli}. In general, one can distinguish at least three different methods of addressing this problem.

First, the wave number $k$-dependence of the turbulent power spectrum $G(k)$ can be specified to derive analytical approximations for $D_{\mu\mu}$. The quasi-linear theory (QLT) derived by \citet{jok66:qlt} has been the standard theory, until it was realized that QLT is not only inaccurate but, in fact, invalid for some scenarios. For the example of isotropic turbulence, it has been known from both qualitative arguments \citep{fis74:iso,bie88:pit} and detailed calculations \citep{tau06:sta} that QLT cannot properly describe pitch-angle scattering, because it neglects $90^\circ$~scattering and leads to infinitely large mean-free paths. This problem was remedied by the application of the second-order QLT \citep[SOQLT, see][]{sha05:soq,tau08:soq}, which considers deviations from the unperturbed spiral orbits that were assumed in QLT.

Second, to allow for more complex turbulence properties and to validate the permissibility of the analytical perturbation theories, one can resort to test particle simulations in specified turbulent magnetic fields. By tracing particle trajectories, the mean square displacements and the associated diffusion parameters can be obtained. On the one hand, there have been successful attempts to confirm quasi-linear results \citep{qin09:dmm}. On the other, such simulations have been employed to investigate the general behavior of pitch-angle scattering as described by the direct summation of multiple particle deflections \citep{lem09:sto}. Further examples of application of this method include the studies of the effects of structured \citep{Laitinen-etal-2012} and balanced turbulence \citep{Laitinen-etal-2013} on the particle transport, and consideration of inhomogeneous magnetic background fields \citep{Tautz-etal-2011,Kelly-etal-2012}.

Third, rather than entirely specifying the turbulent magnetic fields, one can perform direct numerical simulations to compute solutions to the magnetohydrodynamic equations, while the test-particle trajectories are still integrated as in the previous method. Such computations \citep[see, e.g.,][]{Beresnyak-etal-2011, Spanier-Wisniewski-2011, Wisniewski-etal-2012} do not require assumptions regarding the turbulence spectrum that is seen by the energetic particles. They are, however, limited regarding the extent of the inertial range of the turbulence spectrum, owing to computational constraints. In the present study we, therefore, restrict ourselves to the first and second approaches.

We undertake a systematic comparison between analytical predictions of the Fokker-Planck coefficient of pitch-angle scattering and numerical simulations that are based on a Monte-Carlo code developed by one of us \citep{tau10:pad}. In Sect.~\ref{padian}, the Monte-Carlo code \textsc{Padian} is introduced, which is used for all numerical simulations. In Sects.~\ref{pitch} and \ref{fokker}, basic properties of pitch-angle scattering and of the Fokker-Planck coefficient are presented, respectively. In Sect.~\ref{results}, results from the numerical simulations are compared to estimations obtained from analytical scattering theories. Sect.~\ref{summ} provides a summary and a discussion of the results.

\section{\textsc{Padian} simulation code}\label{padian}

For the numerical simulations, a Monte-Carlo code was used to compute the parallel diffusion coefficient of energetic particles for the turbulence model described above. A general description of the code and the underlying numerical techniques can be found elsewhere \citep{tau10:pad,tau13:num}. Specifically, the isotropic and the slab turbulence models were employed, which are defined via the wave vector of the (Fourier transformed) turbulent magnetic field with random orientation and aligned with the direction of the background magnetic field (i.e., the $z$ axis), respectively. The corresponding generation of turbulent magnetic fields proceeded as \citep{tau13:num}
\be\label{eq:dB}
\delta\f B(\f r,t)=\sum_{n=1}^{N_m}\f e'_\perp A(k_n)\cos\left[k_nz'+\beta_n\right],
\ee
where the wavenumbers $k_n$ are distributed logarithmically in the interval $k_{\text{min}} \leqslant k_n \leqslant k_{\text{max}}$, and $\beta$ is a random phase angle.

The slab turbulence model is motived by Mariner~2 measurements, indicating that the solar wind is dominated by outward propagating Alfv\'enic turbulence \citep{bel71:alf,tau13:pal}. Together with a second, two-dimensional contribution, this gave rise to the composite turbulence model \citep{bie96:two,mat90:mal,rau13:mal}. In numerical simulations, in contrast, Alfv\'enic modes exhibit a scale-dependent anisotropy consistent with the Goldreich-Sridhar (\citeyear{sri94:tur},\citeyear{gol95:tur}) model. Nevertheless, the classic slab model remains attractive, especially for analytical investigations, thereby allowing one to isolate specific effects.

For the amplitude and the polarization vector, one has $A(k_n)\propto\sqrt{G(k_n)}$ and $\f e'_\perp\cdot\f e'_z=0$, respectively, with the primed coordinates determined by a rotation matrix with random angles so that $\f k\parallel\ez$ for slab modes and random $\f k$ directions for isotropic modes. From the integration of the Newton-Lorentz equation for the particle motion, various diffusion coefficients can then be calculated by averaging over an ensemble of particles and by determining the mean square displacement. For example, the scattering mean free path in the direction parallel to the background magnetic field can be obtained as $\lambda\pa=(3/v)\,\langle(\De z)^2\rangle/(2t)$ for large times (cf. Sec.~\ref{pitch}).

For the minimum and maximum wavenumbers included in the turbulence generator, the following considerations apply: (i) the \emph{resonance condition} states that there has to be a wavenumber $k$ so that $R_{\text L}k\approx1$, where $R_{\text L}$ denotes the particle's Larmor radius so that scattering predominantly occurs when a particle can interact with a wave mode over a full gyration cycle; (ii) the \emph{scaling condition} requires that $R_{\text L}\Om_{\text{rel}}t<L_{\text{max}}$, where $\Om_{\text{rel}}=qB/(\gamma mc)$ is the relativistic gyrofrequency and where $L_{\text{max}}\propto1/k_{\text{min}}$ is the maximum extension of the system, which is given by the lowest wavenumber (for which one has $k_{\text{min}}=2\pi/\lambda_{\text{max}}$, thereby proving the argument). In practice, the second condition determines the minimum wavenumber, while the first one determines the maximum wavenumber.

Here, values are chosen as $k_{\text{min}}\ell_0=10^{-4}$ and $k_{\text{max}}\ell_0=10^4$, where $\ell_0$ is the turbulence bend-over scale (see Appendix~\ref{app:analyt}). The sum in Eq.~\eqref{eq:dB} extends over $N_m=512$ wave modes, which is sufficient \citep{tau13:num} and yet saves computation time. Furthermore, the maximum simulation time is determined as $vt/\ell_0=10^1$ for low particle energies and $10^2$ for high particle energies.

\section{Pitch-angle scattering}\label{pitch}

\begin{figure}[tb]
\centering
\includegraphics[width=\linewidth]{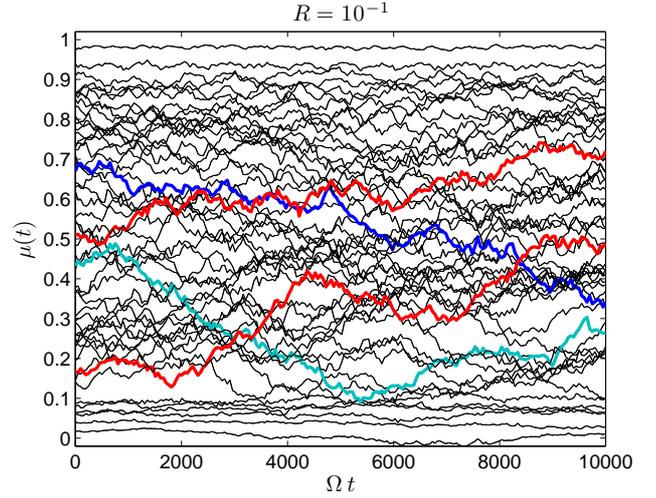}
\caption{(Color online) Pitch-angle cosine, $\mu=v\pa/v$ as a function of the normalized time, $\tau=\Om t$ for a relative slab turbulence strength $\delta B/B_0=10^{-2}$. Four particles with initial pitch angles in the range $0.1\lesssim\mu_0\lesssim0.8$ are highlighted to guide the eye.}
\label{ab:pitchangleplot_B-2_R-1}
\end{figure}

Perhaps the most important transport process of high-energy particles is represented by pitch-angle scattering, i.e., by stochastic variations in $\mu=\cos\angle(\f v,\bo)=v\pa/v$ with range $[-1,1]$, where $\bo=B_0\ez$ is the mean magnetic field and $\f v$ is the particle velocity. This process is related to diffusion along the mean magnetic field, which is described by the parallel diffusion coefficient, $\kappa\pa$, or the parallel mean free path, $\lambda\pa=(3/v)\kappa\pa$, which are also related to the cosmic ray anisotropy \citep{sch89:cr1,sha09:hil}.

The time evolution of the pitch angle is shown in Fig.~\ref{ab:pitchangleplot_B-2_R-1} for a sample of typical single-particle trajectories (without any averaging process). It is indeed confirmed that particles with $\mu\approx\{0,\pm1\}$ almost retain their original pitch angle. However, scattering through $90^\circ$ can occur, a fact that is \emph{not} included in QLT. An introduction to analytical transport theories can be found, e.g., in \citet{rs:rays} and \citet{sha09:nli}.

\subsection{General remarks}\label{pitch:gen}

The usual definition of pitch-angle scattering can be found in the so-called Taylor-Green-Kubo (TGK) formalism \citep{tay22:dif,gre51:bro,kub57:tgk,sha11:tgk} as
\bs\label{eq:tgk}
\begin{align}
\dm(\mu)&=\int_0^\infty\df t\,\m{\dot\mu(t)\dot\mu(0)} \label{eq:tgk_a}\\
&=\frac{1}{2}\,\dd t\,\m{\left(\De\mu(t)\right)^2}, \label{eq:tgk_b}
\end{align}
\es
where the second line employs the definition $\De\mu(t)=\mu(t)-\mu(0)$. It can be shown \citep[e.g.,][]{sha11:tgk} that both versions agree with each other, if $t$ is high enough that the expression becomes asymptotically time-independent.

The combination of diffusion \citep{fic55:dif} and random walk \citep{cha43:sto} motivated the usual definition of the diffusion in terms of the mean-square displacement \citep[e.g.,][]{tau12:nov} $\kappa=\langle(\De x)^2\rangle/(2t)$, which can also be used for a third expression for \dm, namely
\be\label{eq:dmdif}
\dm(\mu)=\frac{1}{2t}\m{\left(\De\mu(t)\right)^2},
\ee
which is again valid if $t$ is high enough. However, the formal limit $t\to\infty$ is forbidden since $\abs{\De\mu}$ cannot exceed a value of~2. For high enough times, $\dm$ will always be dominated by the $1/t$ dependence, independent of the choice of the formula. Therefore, a meaningful, time-independent value for $\dm$ can be obtained if and only if: (i) $t$ is long enough that the initial conditions become insignificant; (ii) $t$ is short enough that the behavior of $\dm$ is not already dominated by the $1/t$ proportionality. This matter is further investigated in the second paper of this series \citep{tau13:pi2}.

\begin{figure}[tb]
\centering
\includegraphics[bb=-10 -10 320 258,width=\linewidth]{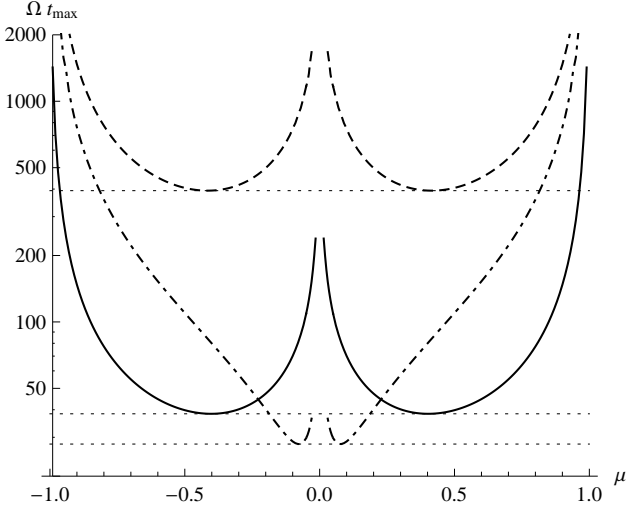}
\caption{Critical time, $t_{\text{max}}$, as obtained from the numerical solution of Eq.~\eqref{eq:Dm2} for particles with different pitch angles. The solid and dot-dashed lines show the cases of particles with rigidity values chosen as $R=1$ and $R=10$, respectively, with a relative turbulence strength of $\delta B/B_0=10^{-1}$. For the dashed line, parameters are $R=1$ and $\delta B/B_0=10^{-1.5}$. The dotted lines show the minimum of the critical time for all pitch angles.}
\label{ab:DmTcrit}
\end{figure}

\subsection{Estimation of the critical time}\label{pitch:tcrit}

Based on the preceding paragraph, we now calculate the critical time, $t_{\text{max}}$, by using QLT. This gives us an estimates for the time range, during which we can expect Eqs.~\eqref{eq:tgk_b} and \eqref{eq:dmdif} to be valid.

Following the derivation of the quasi-linear Fokker-Planck coefficient \citep[see][]{sha05:soq}, the pitch-angle displacement, $\De\mu=\mu(t)-\mu_0$, can be expressed as
\be
\De\mu(t)=\frac{\Om}{vB_0}\int_0^t\df t'\,\left[v_x(t')\,\delta B_y-v_y(t')\,\delta B_x\right],
\ee
where
\bs
\begin{align}
v_x&=v\sqrt{1-\mu^2}\cos\!\left(\Ph_0-\Om t\right)\\
v_y&=v\sqrt{1-\mu^2}\sin\!\left(\Ph_0-\Om t\right)
\end{align}
\es
denote the unperturbed, quasi-linear, orbits with $\Om$ the gyrofrequency and $\Ph_0$ the initial gyro phase.

Squaring and taking the ensemble average yields
\begin{align}
\m{\left(\De\mu(t)\right)^2}&=\frac{\Om^2(1-\mu^2)}{B_0^2}\int_0^t\df t'\int_0^t\df t''\nonumber\\
&\times\cos\left[\Om\left(t'-t''\right)\right]\m{\delta B_x(t')\,\delta B_x(t'')}. \label{eq:tmp}
\end{align}
After taking the Fourier transform, the magnetic correlation tensor element $\mathsf P_{xx}(\f k)$ can be inserted.

\begin{figure}[tb]
\centering
\includegraphics[width=\linewidth]{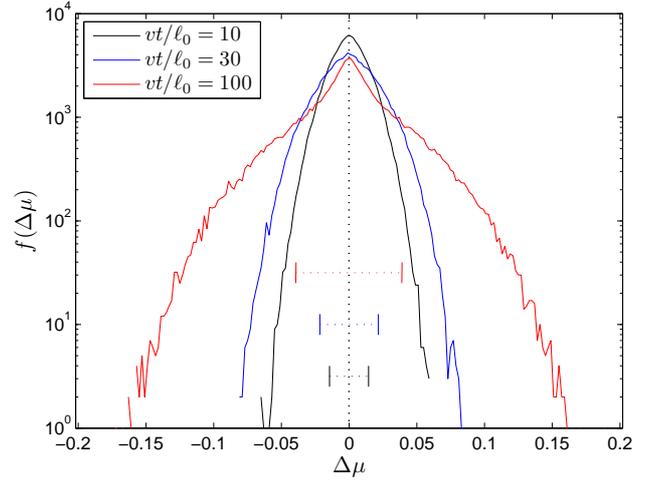}
\caption{(Color online) Distribution function $f(\De\mu)$ of the relative pitch-angle displacement, $\De\mu=\mu(t)-\mu_0$, for three different times. The horizontal dotted bars show the corresponding standard deviations. Additionally, the vertical dotted line shows the mean (which remains almost unchanged).}
\label{ab:pitch_slab_dmDistri}
\end{figure}

For instance, by assuming slab geometry and setting $\tau=\Om t$, the result can finally be expressed as \citep[cf.][]{sha05:soq}
\be\label{eq:Dm2}
\m{\left(\De\mu(\tau)\right)^2}=4\pi\left(1-\mu^2\right)\left(\frac{\delta B}{B_0}\right)^2\int_0^\infty\df k\;G(k)\left[\mathcal K_++\mathcal K_-\right],
\ee
where the spectrum $G(k)$ from Eq.~\eqref{eq:spect} has been used. In what follows, normalized variables are used as $R=\gamma v/(\Om\ell_0)$, and $x=\ell_0k$; the resonance functions can then be expressed as
\be
\mathcal K_\pm=\frac{1-\cos\left[\left(xR\mu\pm1\right)\tau\right]}{\left(xR\mu\pm1\right)^2}.
\ee
By requiring that $\lvert\De\mu_{\text{max}}\rvert\leqslant1$, the critical time $t_{\text{max}}$ can be obtained by (numerically) solving Eq.~\eqref{eq:Dm2} for the time. The result is shown in Fig.~\ref{ab:DmTcrit} and confirms the estimation that, for stronger turbulence, the increased scattering causes particles to reach the limiting pitch angle values earlier.

For other turbulence geometries---especially isotropic turbulence---the same analysis is possible in principle, although the required calculations become somewhat unwieldy \citep[cf.][]{tau08:soq,tau10:soq}.

\subsection{Pitch-angle distribution}

In this paragraph, it is demonstrated that particles stay in their original pitch-angle regime for all times. This is true even if, as stated above, the Fokker-Planck coefficient of pitch-angle scattering tends to zero for times $t>t_{\text{max}}$.

In Fig.~\ref{ab:pitch_slab_dmDistri}, the distribution function of all particles sorted for their pitch-angle displacement, $\De\mu$, is shown. As time increases, particles begin to deviate from their original pitch angles. Nevertheless (note the logarithmic scaling of the vertical axis!), the distribution remains extremely small with its width growing linearly with time as $10^{-5}\Om t$.

\begin{figure}[tb]
\centering
\includegraphics[width=\linewidth]{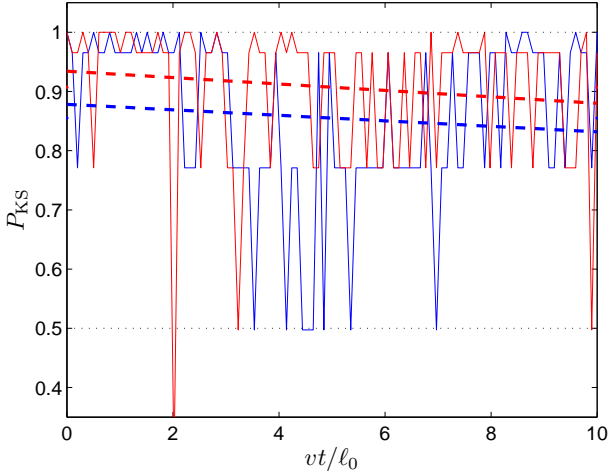}
\caption{(Color online) Time evolution of the KS test statistic (solid lines) for the agreement between the pitch-angle distribution, $f(\mu,t)$, and the initial distribution, $f(\mu_0,0)$ (which are obtained from uniform random deviates) together with linear fits (dashed lines). The red and blue lines correspond to the cases of low and intermediate turbulence strength, respectively.}
\label{ab:pitch_slab_KS}
\end{figure}

In Fig.~\ref{ab:pitch_slab_KS}, the time evolution of a Kolmogorov-Smirnov (KS) test statistic is shown, which is obtained from the comparison of the pitch-angle distribution, $f(\mu,t)$, with the initial pitch-angle distribution, $f(\mu_0,0)$. Because the latter is obtained from a uniform random deviate, the usual result of an isotropized distribution due to pitch-angle diffusion leads to the requirement that the pitch-angle distribution be uniform. A linear fit of the otherwise relatively volatile KS statistic shows that, on average, the $P$~value (i.e., the probability that the given distribution agrees with the assumed one) of the KS~test is well above 80\%; this confirms that the pitch-angle distribution remains compatible with the initial distribution.

This result serves as a second indicator that an initially homogeneous pitch-angle distribution is retained. As a side note, it should be mentioned that the pitch-angle distribution is not \emph{precisely} homogeneous, i.e., the comparison of the pitch-angle distribution with a \emph{flat} distribution yields a reduced KS test statistic as opposed to the comparison of $f(\mu,t)$ with $f(\mu_0)$.

\begin{table}[b]
\centering
\begin{tabular}{llllll}\hline\hline
$\delta B/B_0$		& Rigidity		& Theory 	& $\chi^2$ value 	& $Q$ value			& $vt_{\text{max}}/\ell_0$\\\hline
$10^{-2}$			& $10^{-2}$		& QLT		& $24.47$			& $0.9747$			& $1$\\
$10^{-2}$			& $10^{-2}$		& SOQLT		& $24.33$			& $0.9759$			& $1$\\
$10^{-2}$			& $10^{-1}$		& QLT		& $6.55$			& $\approx1$		& $10$\\		
$10^{-2}$			& $10^{-1}$		& SOQLT		& $6.43$			& $\approx1$		& $10$\\		
$10^{-2}$			& $1$			& QLT		& $2.53$			& $\approx1$		& $100$\\
$10^{-2}$			& $1$			& SOQLT		& $28.81$			& $0.9057$			& $100$\\
$10^{-1.5}$			& $10^{-2}$		& QLT		& $16.74$			& $0.9956$			& $1$\\
$10^{-1.5}$			& $10^{-2}$		& SOQLT		& $16.60$			& $0.9960$			& $1$\\
$10^{-1.5}$			& $10^{-1}$		& QLT		& $10.23$			& $\approx1$		& $10$\\		
$10^{-1.5}$			& $10^{-1}$		& SOQLT		& $10.37$			& $\approx1$		& $10$\\		
$10^{-1.5}$			& $1$			& QLT		& $13.55$			& $\approx1$		& $100$\\		
$10^{-1.5}$			& $1$			& SOQLT		& $27.71$			& $0.9296$			& $100$\\
$10^{-1.5}$			& $10$			& QLT		& $2.24$			& $\approx1$		& $100$\\
\hline\hline
\end{tabular}
\caption{Overall agreement between numerical results for the pitch-angle Fokker-Planck coefficient, \dm, and both quasi-linear (QLT) and second-order quasi-linear (SOQLT) analytical results as obtained from a chi-square test.}
\label{ta:param}
\end{table}

\section{Fokker-Planck coefficient}\label{fokker}

In this section, some of the intricacies connected to the numerical implementation of pitch-angle scattering is discussed. For an overview of analytical calculations, the reader is referred to Appendix~\ref{app:analyt}, where both quasi-linear and nonlinear results are summarized.

\begin{figure}[tb]
\centering
\includegraphics[width=\linewidth]{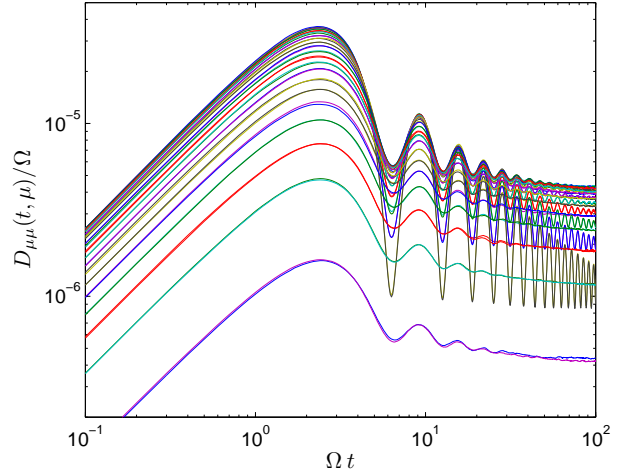}
\caption{(Color online) Running pitch-angle Fokker-Planck coefficient, $\dm(t)$, for various values of the initial pitch-angle cosine, $\mu_0$, which is represented by the different colors. For the numerical evaluation, Eq.~\eqref{eq:dmdif} has been used, together with the procedure described thereafter. Other parameters are identical to those in Fig.~\ref{ab:pitchangleplot_B-2_R-1}.}
\label{ab:Tdmu_slab_B-2_R-1}
\end{figure}

While the calculation of the mean free path values is straightforward, this is somewhat different for the Fokker-Planck coefficient(s). Especially $\dm$ has to be a function of time but, at the same time, depends on the $\mu$ values. It has to be stressed that pitch-angle scattering is unique in that, unlike for normal spatial diffusion, the coefficient \emph{depends} on the variable from which the mean square displacement is obtained.

Therefore, several options are possible, two of which will be described here.

\begin{itemize}
\item Particles are binned once and for all to $\mu$ slots according to their \emph{initial} pitch-angle, $\mu(0)\equiv\mu_0$. While at first glance this procedure seems to be justified by the fact that Eq.~\eqref{eq:dmdif} is symmetric in $\mu(t)$ and $\mu_0$, it has the crucial drawback of conserving the initial conditions. However, a diffusive process is defined as a process that has become independent of the initial condition; therefore, this approach has to be discarded.

\item In an alternative approach, as opposed to the one described above, particles are binned separately at each point in time to $\mu$ slots. Average values are then taken for all particles in a given $\mu$ slot, thereby obtaining $\dm(\mu,t)$ as a function of pitch angle and time. In principle, both Eqs.~\eqref{eq:tgk_b} and \eqref{eq:dmdif} should be interchangeable. In practice, however, it turns out that the derivative is considerably more volatile, while Eq.~\eqref{eq:dmdif} yields reasonable results even for a moderate number of particles.
\end{itemize}

\begin{figure}[tb]
\centering
\vspace{2ex}
\includegraphics[width=\linewidth]{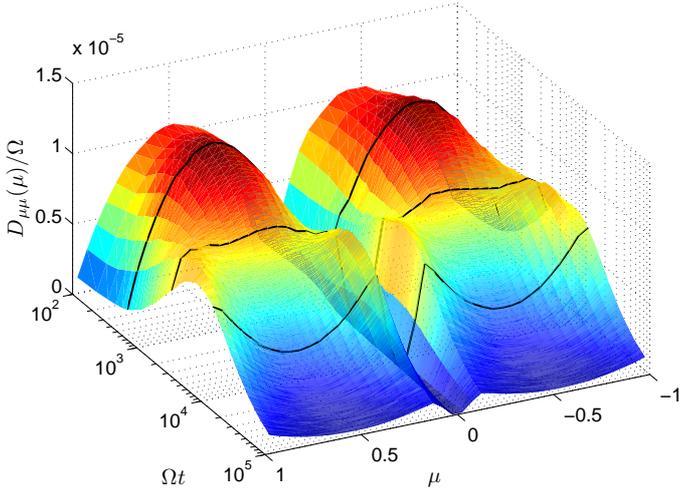}
\caption{(Color online) Running Fokker-Planck coefficient of pitch-angle scattering, $\dm(\mu,t)$ for particles width rigidity $R=10^{-2}$ in moderate turbulence strength, $\delta B/B_0=10^{-1.5}$. The black solid lines illustrate $\dm(\mu)$ at specific times.}
\label{ab:dmumu_slab_3D}
\end{figure}

\begin{figure}[bt]
\centering
\includegraphics[width=\linewidth]{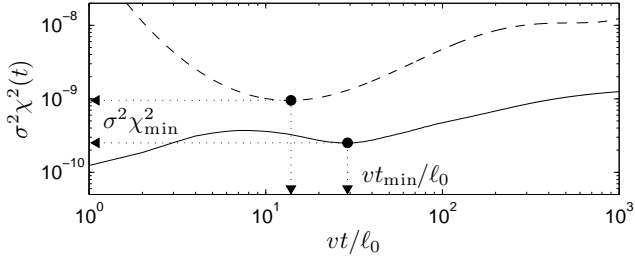}
\caption{Time evolution of the $\chi^2$ value associated with the running Fokker-Planck coefficient of pitch-angle scattering as shown in Fig.~\ref{ab:dmumu_slab_3D}. The solid and dashed lines show the cases of $R=10^{-2}$ and $R=10^{-1}$, respectively, with $\delta B/B_0=10^{-1.5}$ for both cases. The $\chi^2$ value is obtained from the comparison with Eq.~\eqref{eq:dmQLT}.}
\label{ab:dmumu_slab_KS2}
\end{figure}

The second approach is illustrated in Fig.~\ref{ab:Tdmu_slab_B-2_R-1} for various pitch angles. By evaluating Eq.~\eqref{eq:dmdif}, a Fokker-Planck coefficient is obtained that depends both on the time and on the pitch-angle cosine. Plotted as a function of time, the transition to the diffusive regime can be seen for times $t\simeq5\Om^{-1}$. For times $t\geqslant50\Om^{-1}$, it is admissible to take asymptotic values for the Fokker-Planck coefficient as a function of the pitch-angle cosine, i.e., $\dm(\mu)$.

In Fig.~\ref{ab:dmumu_slab_3D}, the time evolution of the pitch-angle Fokker-Planck coefficient is illustrated. The typical double-hump structure known from analytical theories (cf. Appendix~\ref{app:analyt}) is exhibited; for later times, in contrast, the characteristic $1/t$ dependence shown in Eq.~\eqref{eq:tgk_b} dominates. Additionally, the shape of $\dm$ is strongly modified, thereby resulting in a drastically increased $\chi^2$ for the comparison between simulation and analytical theory as shown in Fig.~\ref{ab:dmumu_slab_KS2}.

\section{Comparison with analytical results}\label{results}

In this section, the numerical results for the Fokker-Planck coefficient will be compared to analytical results listed in Appendix~\ref{app:analyt}. Error bars are obtained from the comparison of different turbulence realizations and different initial particle positions \citep[see][]{tau10:pad}. In addition, it has to be stressed again that, according to Fig.~\ref{ab:dmumu_slab_3D}, the correct time point has to be chosen for the evaluation of \dm.

\subsection{Slab turbulence}

\begin{figure}[tb]
\centering
\includegraphics[width=\linewidth]{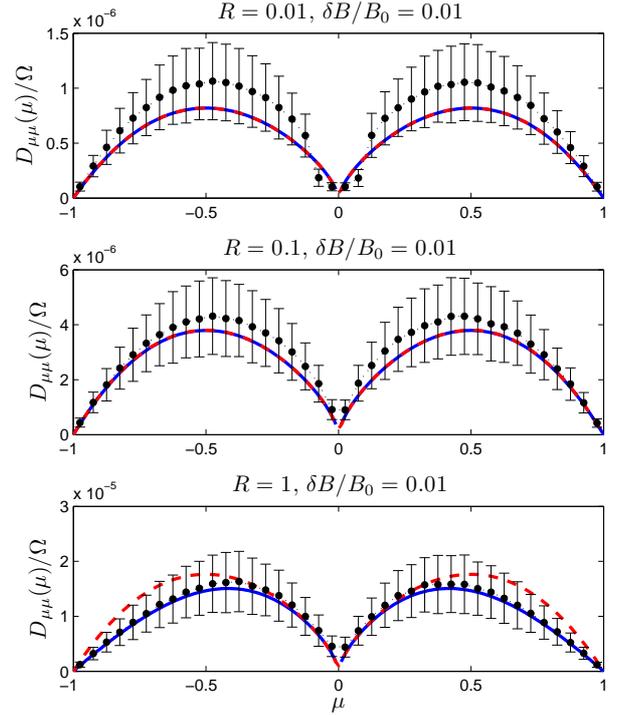}
\caption{(Color online) Numerically calculated pitch-angle Fokker-Planck coefficient, \dm, as a function of the pitch-angle cosine, $\mu$, for three different values for the normalized rigidity, $R$. For comparison, the analytical results from QLT and SOQLT are shown as solid blue and dashed red lines, respectively. The relative turbulence strength is chosen as $\delta B/B_0=10^{-2}$.}
\label{ab:dmumu_slab_B-2_R}
\end{figure}

In Fig.~\ref{ab:Tdmu_slab_B-2_R-1}, the pitch-angle Fokker-Planck coefficient is shown as a function of the normalized simulation time, $\tau=\Om t$. After the initial free-streaming phase, most values become (almost) constant, while the initially higher values still oscillate. At $\tau=10^2$, the final, diffusive values for $\dm$ are taken that are used in the following sections. It should be noted, however, that $\dm$ is slightly decreasing (with approximately $\propto\tau^{-0.14}$) so that the values for $\dm$ are somewhat overestimated, in agreement with the results shown below.

In the following, the two cases of low and intermediate turbulence strengths are discussed.

\subsubsection{Low turbulence strength}

For the ratio of the turbulent and background magnetic fields, a value of $\delta B/B_0=10^{-2}$ is chosen so that the ratio of the magnetic field energies is $(\delta B/B_0)^2=10^{-4}$. The following results were found (see Fig.~\ref{ab:dmumu_slab_B-2_R}):

\begin{itemize}
\item For small and intermediate rigidities ranging from $R=10^{-2}$ to $10^{-1}$, an excellent agreement between numerical and analytical results can be found. However, QLT and SOQLT are almost indistinguishable. For example, a chi-square test yields values of $\chi^2=6.551$ and $6.432$ at $R=10^{-1}$ for the comparison to QLT and SOQLT, respectively, thereby revealing that the agreement with SOQLT is slightly better. However, the difference is marginal and might have occurred purely by serendipity.

\item For high rigidities such as $R=1$ and $R=10$, QLT and SOQLT differ more. A chi-square test yields values of $\chi^2=2.531$ and $28.81$ at $R=1$ for the comparison to QLT and SOQLT, respectively, thereby revealing that the agreement with QLT is \emph{significantly} better. However, it should be noted that the approximation used for the SOQLT values becomes invalid if $R$ is too large.
\end{itemize}

\subsubsection{Intermediate turbulence strength}

\begin{figure}[tb]
\centering
\includegraphics[width=\linewidth]{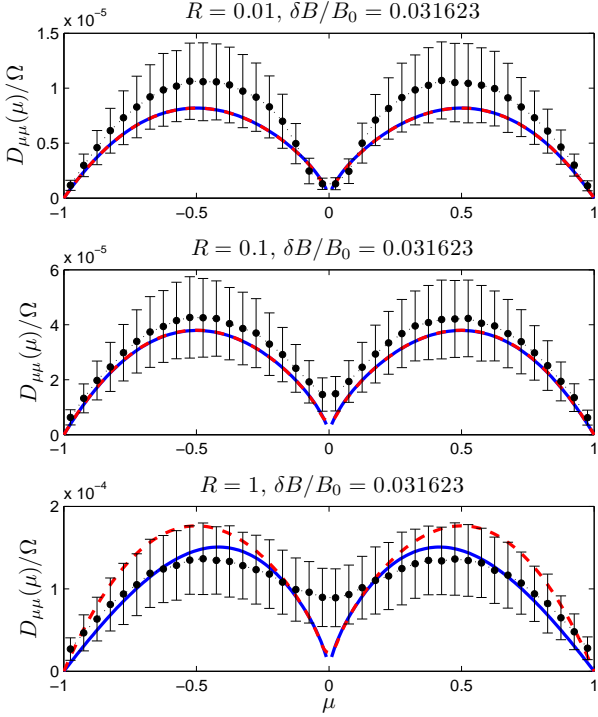}
\caption{(Color online) Same as Fig.~\ref{ab:dmumu_slab_B-2_R}, only that now the relative turbulence strength is chosen as $\delta B/B_0=10^{-1.5}\approx0.0316$.}
\label{ab:dmumu_slab_B-1_R}
\end{figure}

Here, $\delta B/B_0=10^{-1.5}\approx0.0316$ is chosen so that the ratio of the magnetic field energies is $10^{-3}$. The following results were found (see Fig.~\ref{ab:dmumu_slab_B-1_R}):

\begin{itemize}
\item For small and intermediate rigidities, the simulation results agree equally well with both QLT and SOQLT, to the same level of significance as was found in the previous section.

\item For high rigidities, it is shown that both QLT and SOQLT severely underestimate $90^\circ$ scattering, even though SOQLT was designed explicitly to remedy this shortcoming of previous, quasi-linear results. Accordingly, the chi-square test yields slightly higher values of $\chi^2=13.55$ and $14.05$ at $R=1$ for QLT and SOQLT, respectively, which again shows that QLT agrees slightly better with the numerical values.

\item Additionally, it is remarkable that, for $R=10$, the overall best agreement has been found as expressed by the low value $\chi^2=2.24$.
\end{itemize}

In general, it has to be noted (cf. Table~\ref{ta:param}) that the agreement between analytical and numerical results depends on the maximum simulation time. For $t\neq t_{\text{max}}$, less agreement is found.

\subsection{Isotropic turbulence}

\begin{figure}[tb]
\centering
\includegraphics[width=\linewidth]{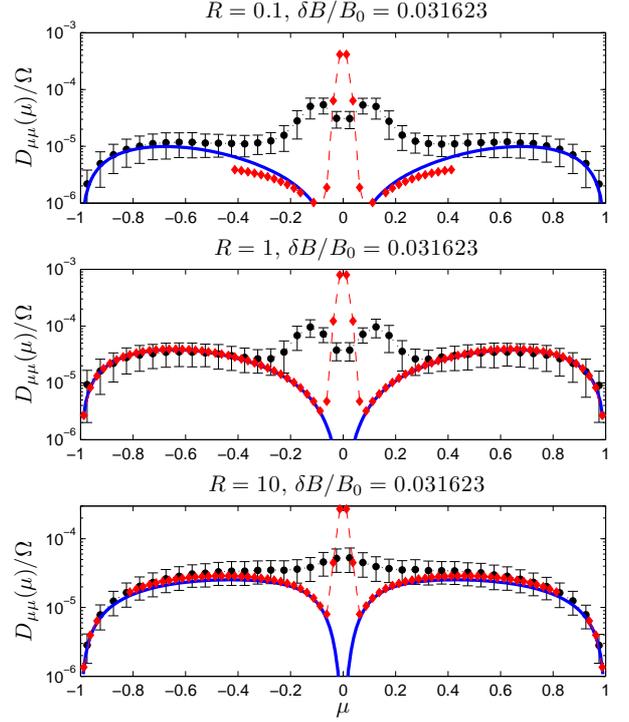}
\caption{(Color online) Same as Fig.~\ref{ab:dmumu_slab_B-1_R}, only that now the case of isotropic turbulence is shown. Due to the complexity of the SOQLT, the evaluation is given at some discrete points instead of as a continuous function.}
\label{ab:dmumu_iso_B-1_R}
\end{figure}

For isotropic turbulence, the numerical Fokker-Planck coefficient is shown in Fig.~\ref{ab:dmumu_iso_B-1_R}. The comparison especially with SOQLT had to be done for higher rigidities than for slab geometry simply because the numerical evaluation of Eq.~\eqref{eq:DmIso} is extremely protracted for low rigidities and/or low turbulence strengths.

While the agreement between theory and simulation is generally good at pitch-angles well off $90^\circ$, it is revealed that $90^\circ$ degree scattering is not equally well described either by QLT or by SOQLT. Therefore, the agreement of the parallel mean free path with simulations \citep[cf.][]{tau08:soq,tau10:soq} has to be attributed to the fact that, at $\mu=0$, SOQLT overestimates \dm, thereby compensating for values that are too low at $0.3\lesssim\mu\leqslant0$.

\section{Summary and conclusion}\label{summ}

In this paper, random variations in the pitch-angle of charged particles that move in a turbulent magnetic field have been investigated. In the astrophysical theater, this situation is realized by cosmic rays and solar particle events, both of which experience continuous deflections either in the interstellar turbulence or in the solar-wind induced turbulence. In both cases, the random component superposes a mean magnetic field---e.g., the galactic magnetic field or the Parker-spiral solar magnetic field---which gives rise to a preferred direction when investigating the scattering processes. This motivates the transformation to a coordinate system in which the pitch angle is taken to be a basic variable of the particle motion.

The process can be analyzed by means of analytic calculations and numerical Monte-Carlo simulations, which are methods based on the kinetic Vlasov theory and the integration of the equation of motion for a large number of test particles, respectively. Whereas the use of a Fokker-Planck approach to determining the pitch-angle scattering and, based thereupon, the parallel mean-free path, has been established decades ago, it has been difficult to reproduce the process using test-particle simulations. The reason is not only that, owing to the required pitch-angle resolution of the Fokker-Planck coefficient, a large number of particles are required, but that, additionally, the underlying algorithm is not entirely clear. While there are previous simulations \citep{qin09:dmm} that confirm quasi-linear results, here we have shown that the time dependence of the Fokker-Planck coefficient cannot be neglected.

Instead, one general result that has been found is the following: because the pitch-angle cosine, $\mu$, is confined to $\mu\in[-1,1]$, pitch-angle scattering as based on Eqs~\eqref{eq:tgk_b} and \eqref{eq:dmdif} is not a process that can be described for asymptotically long times. Instead, the proper time has to be found where (i) $\dm$ is no longer dominated by the initial conditions; but (ii) $\dm$ is not yet dominated by the $1/t$ dependence because $\De\mu$ cannot grow any further. However, with the additional constraint that the turbulence strength is not too high, there always seems to be a time period where excellent agreement with the analytical results can be obtained.

The important question that has to be raised, therefore, is the validity of the generally accepted formulae for the Fokker-Planck coefficient. In order to account for the deviations found in the present paper, in a second paper the underlying theoretical basis of the pitch-angle Fokker-Planck coefficient as based on the diffusion equation will be revisited.

\begin{acknowledgements}
RCT thanks Gary Zank, Fathallah Alouani-Bibi, and Andreas Shalchi for useful discussions on the subject of pitch-angle scattering.
\end{acknowledgements}

\appendix
\section{Analytical pitch-angle scattering}\label{app:analyt}

Analytically, the Fokker-Planck coefficient of pitch-angle scattering can be evaluated, e.g., using quasilinear theory \citep{jok66:qlt} and nonlinear extensions \citep[see][for an overview]{sha09:nli}. In any case, the evaluation is based on the TGK formalism by using Eq.~\eqref{eq:tgk_a}.

\subsection{Slab turbulence}

For slab turbulence, the result is \citep[e.g.,][]{qin09:dmm}
\be
\dm^{\text{QLT}}(\mu)=\frac{2\pi^2v\left(1-\mu^2\right)}{\abs\mu\Rl^2}\left(\frac{\delta B}{B_0}\right)^2G_{\text{slab}}\left(k\pa=\frac{1}{\abs\mu\Rl}\right).
\ee
For the turbulence power spectrum, $G(k)$, a kappa-type function is used \citep{sha09:flr}:
\be\label{eq:spect}
G(k)=\frac{C}{2\sqrt\pi}\,\frac{\Ga(s/2)}{\Ga\bigl((s-1)/2\bigr)}\frac{\abs{\ell_0k}^q}{\left[1+\left(\ell_0k\right)^2\right]^{s/2}},
\ee
with usually $q=0$ for simplicity. The turbulence bend-over scale, $\ell_0\approx0.03$\,AU, reflects the transition from the energy range $G(k)\propto k^q$ to the Kolmogorov-type inertial range, where $G(k)\propto k^{-s}$ with $s=5/3$ for large wavenumbers \citep{kol91:tur,bru05:sol}. The factor $C$ depends on the assumed geometry and is given as $C=1/(2\pi)$ for slab turbulence, where $\delta\f B(\f r)=\delta\f B(z)$, and $C=4$ for isotropic turbulence.

In the normalized variables $R$ and $\tau$, the Fokker-Planck coefficient of pitch-angle scattering can be expressed as
\be\label{eq:dmQLT}
\dm^{\text{QLT}}(\mu)=\frac{2\pi^2}{\abs\mu R}\left(1-\mu^2\right)\left(\frac{\delta B}{B_0}\right)^2G_{\text{slab}}\left(k\pa=\frac{1}{\abs\mu R}\right).
\ee
The typical $(1-\mu^2)$ dependence reflects the fact that particles with pitch angles close to $0^\circ$ and $180^\circ$ are considerably less scattered than particles with intermediate pitch angles; in addition, the rightmost factor in Eq.~\eqref{eq:dmQLT} approximately gives $\lvert\mu\rvert^{2/3}$, thereby suppressing $90^\circ$ scattering.

A nonlinear theory developed especially to enhance pitch-angle scattering through $90^\circ$ \citep{sha05:soq,sha09:nli} yields the formula
\begin{align}
\dm&=\frac{1}{8s\sqrt\pi}\,\frac{\Ga(s/2)}{\Ga\bigl((s-1)/2\bigr)}\left(1-\mu^2\right)C(s)\,\frac{R^{2-s}v}{\ell_0}\,\frac{\delta B}{B_0}\nonumber\\
&\times\sum_{n=\pm1}\text{sgn}\left(\frac{\delta B}{B_0}+n\abs\mu\right)\abs{\abs\mu+n\,\frac{\delta B}{B_0}}^s.
\end{align}
where $R$ and $v$ are the normalized rigidity and the particle speed, respectively.

\subsection{Isotropic turbulence}

For isotropic turbulence, the analytical theory of pitch-angle scattering is considerably more difficult to solve \citep{tau06:sta,tau08:soq}. The general form of the Fokker-Planck coefficient for pitch-angle scattering\footnote{Here we corrected for the additional factor of $\pi$ that was erroneously present in Eq.~(3) of \citet{tau08:soq}.} reads as
\begin{align}
\dm&=2\left(1-\mu^2\right)\Om^2\left(\frac{\delta B}{B_0}\right)^2\int_0^1\df\eta\int_0^\infty\df kG(k)\nonumber\\
&\times\usum\mathcal R_n(k,\eta)\left[\eta^2{J'_n}^2(w)+\frac{n^2}{w^2}\,J_n^2(w)\right], \label{eq:DmIso}
\end{align}
with $J_n(w)$ the Bessel function of the first kind of order $n$ and $w=(kv/\Om)\sqrt{(1-\mu^2)(1-\eta^2)}$. Additionally, $\eta=\cos\angle(\f k,\bo)$ is the wave vector polar angle. The resonance function can be expressed as
\bs
\be
\mathcal R_n(k,\eta)=\pi\delta\left(kv\mu\eta+n\Om\right)
\ee
for QLT and
\be
\mathcal R_n(k,\eta)=\sqrt{\frac{\pi}{2}}\left(\xi k\eta\right)^{-1}\exp\left[-\frac{\left(kv\mu\eta+n\Om\right)^2}{2\left(k\eta\xi\right)^2}\right]
\ee
\es
for SOQLT, where $\xi=v^2(\delta B/B_0)^2/3$. Using QLT, Eq.~\eqref{eq:DmIso} can be simplified further \citep[see][]{tau06:sta}, whereas, for SOQLT, the two integrals and the infinite sum have to be evaluated numerically.

Alternatively, the formulation of \citet{tau10:soq} can be used for the case of SOQLT, where the infinite sum over Bessel functions was reduced to a closed form analytical expression under the assumption of a Cauchy-type resonance function.



\end{document}